\documentclass{INTERSPEECH2023}

\interspeechcameraready

\usepackage{enumitem}
\usepackage{multirow}
\usepackage{makecell}
\usepackage{xhfill}
\usepackage{etoolbox}
\usepackage{tikz}
\usepackage{pgfplots}
\definecolor{plt_tabblue}{RGB}{31, 119, 180}
\definecolor{plt_tabyellow}{RGB}{255, 127, 14}
\definecolor{plt_tabgreen}{RGB}{44, 160, 44}
\definecolor{plt_tabgrey}{RGB}{127, 127, 127}
\usepgfplotslibrary{groupplots}
\usepgfplotslibrary{statistics}
\pgfplotsset{compat=1.16}
\makeatletter
\pgfplotsset{
  grid style={dotted, gray},
  boxplot/lower notch/.initial=\pgfplotsboxplotvalue{median},
  boxplot/upper notch/.initial=\pgfplotsboxplotvalue{median},
  boxplot/notch width/.initial=0.5,
  boxplot/draw/box/.code={%
    \draw[/pgfplots/boxplot/every box/.try]
      (boxplot box cs:\pgfplotsboxplotvalue{lower quartile},0)
      -- (boxplot box cs:\pgfplotsboxplotvalue{lower notch},0)
      -- (boxplot box cs:\pgfplotsboxplotvalue{median},0.5-\pgfplotsboxplotvalue{notch width}/2)
      -- (boxplot box cs:\pgfplotsboxplotvalue{upper notch},0)
      -- (boxplot box cs:\pgfplotsboxplotvalue{upper quartile},0)
      -- (boxplot box cs:\pgfplotsboxplotvalue{upper quartile},1)
      -- (boxplot box cs:\pgfplotsboxplotvalue{upper notch},1)
      -- (boxplot box cs:\pgfplotsboxplotvalue{median},0.5+\pgfplotsboxplotvalue{notch width}/2)
      -- (boxplot box cs:\pgfplotsboxplotvalue{lower notch},1)
      -- (boxplot box cs:\pgfplotsboxplotvalue{lower quartile},1)
      -- cycle
    ;
  },
  boxplot/draw/median/.code={%
    \draw[/pgfplots/boxplot/every median/.try]
        (boxplot box cs:\pgfplotsboxplotvalue{median},0.5-\pgfplotsboxplotvalue{notch width}/3*2)
        --
        (boxplot box cs:\pgfplotsboxplotvalue{median},0.5+\pgfplotsboxplotvalue{notch width}/3*2)
    ;
  }
}
\pgfplotsset{
  /pgfplots/custom legend/.style={
      legend image code/.code={
          \draw [|-|,#1] (0,2mm) -- node[rectangle,minimum size=2.5mm,draw,fill,##1]{}
          (0,7mm);
      }
  }
}

\title{DeepFilterNet: Perceptually Motivated Real-Time Speech Enhancement}
\name{Hendrik Schröter$^1$, Alberto N. Escalante-B.$^2$, Tobias Rosenkranz$^2$, Andreas Maier$^1$}
\address{
  $^1$Friedrich-Alexander-Universit\"at Erlangen-N\"urnberg, Pattern Recognition Lab\\
  $^2$WS Audiology, Research and Development, Erlangen, Germany
}
\email{hendrik.m.schroeter@fau.de}

\newcommand{\C}{\mathbb{C}}

\begin{document}

\maketitle
 
\begin{abstract}
  \vspace{-.2em}
  Multi-frame algorithms for single-channel speech enhancement are able to take advantage from short-time correlations within the speech signal.
  Deep Filtering (DF) was proposed to directly estimate a complex filter in frequency domain to take advantage of these correlations.

  In this work, we present a real-time speech enhancement demo using DeepFilterNet.
  DeepFilterNet's efficiency is enabled by exploiting domain knowledge of speech production and psychoacoustic perception.
  Our model is able to match state-of-the-art speech enhancement benchmarks while achieving a real-time-factor of 0.19 on a single threaded notebook CPU.
  The framework as well as pretrained weights have been published under an open source license.
\end{abstract}
\vspace{-1em}

\section{Introduction}
\vspace{-.2em}
Recently, various speech enhancement models take advantage of properties of psychoacoustic perception.
That is, a speech model consisting of a periodic (voiced) and a noisy component is assumed~\cite{valin2018rnnoise, valin2020perceptually, schroeter2022deepfilternet}.
Further, it can be assumed that the exact phase of the noisy speech component is not relevant. Rather reconstructing the speech envelope at a course frequency resolution is sufficient so that the resulting component sounds \textit{like} the original speech.
For the voiced speech however, reconstructing the phase, or, improving the periodicity is important in low signal-to-noise ratio (SNR) conditions.
Multi-frame (MF) filtering in frequency domain was proposed to take advantage of short-time speech correlations, since speech can be decomposed into a correlated and an interfering component~\cite{huang2011multi}.
Deep filtering (DF) \cite{schroeter2020clcnet, mack2019deep} has been recently used to directly estimate the complex frequency domain filter that is able to enhance the periodic component.
It has been shown, that deep filtering outperforms the mostly used complex ratio mask (CRM)~\cite{schroeter2022deepfilternet} and achieves state-of-the-art results~\cite{schroeter2022deepfilternet2}.

In this work, we summarize our proposed DeepFilterNet framework and show some improved results over previous work~\cite{schroeter2022deepfilternet2}.
Due to its efficiency, we can use the model for real-time noise reduction e.g.~in video-calls.

\section{Deep Filtering}
\vspace{-.2em}
\label{sec:df}
\subsection{Signal Model}
\vspace{-.2em}
\label{ssec:signalmodel}
Let $x(k)$ be a mixture signal
\vspace{-.25em}\begin{equation}
  x(k) = s(k) + z(k)\text{,}
  \label{eq:sigmodel}
  \vspace{-.2em}
\end{equation}
where $s(k)$ is a clean speech signal and $z(k)$ an interfering background noise.
Typically, noise reduction operates in time/frequency domain:
\begin{equation}
  X(t, f) = S(t, f) + Z(t, f)\text{,}
  \vspace{-.2em}
\end{equation}
where $X(t, f)$ is the STFT representation of the time domain signal $x(k)$ and $t$, $f$ are the time and frequency bins.
To simplify the following, we omit the frequency index $f$ since all frequency bins are processed equivalently.
Further, with filter length $N$, we define the noisy multi-frame vector as $\bm{\bar{x}}_{t} \in \C^N$:
\begin{equation}
  \bm{\bar{x}}(t)= [X(t + l), X(t - 1 + l), \dots, X(t -N+1 +l)]^\text{T}\text{\ ,}
  \vspace{-.2em}
\end{equation}
where $l$ is an optional look-ahead parameter.
Due to the look-ahead, the filter will include non-causal taps which introduces additional latency.
And with the complex filter $\bm{\bar{w}}(t) \in \C^N$
\begin{equation}
  \bm{\bar{w}}_\text{DF}(t)= [W_0(t), W_1(t), \dots, W_{N-1}(t)]^\text{T}
  \vspace{-.2em}
\end{equation}
we define the deep filtering as:
\begin{equation}
  \boxed{Y(t) = \bm{\bar{w}}_{\text{DF}}(t)^{\text{H}}\bm{\bar{x}}(t)}\text{\ ,}
  \label{eq:df2}
  \vspace{-.2em}
\end{equation}
where $\circ^\text{H}$ denotes the conjugate transpose operator.
As mentioned above, deep filtering directly estimates the complex filter $\bm{\bar w}_\text{DF}(t)$.
\vspace{-.4em}

\section{DeepFilterNet Framework}
\vspace{-.2em}
\label{sec:framework}
\begin{figure}
  \includegraphics[width=\linewidth,trim=0 .7cm 0 0]{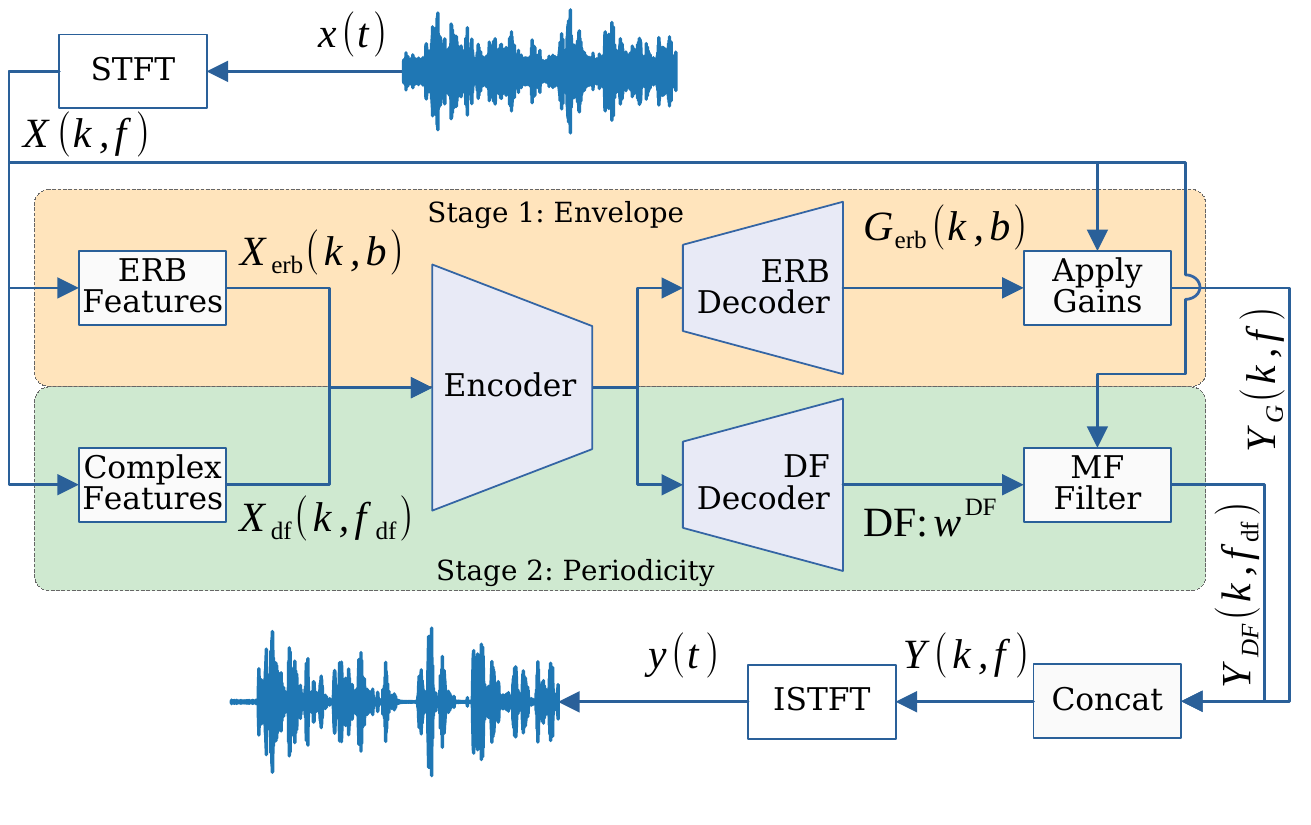}
  \caption{DeepFilterNet Framework.}
  \label{fig:DeepFilterNet}
  \vspace{-.7em}
\end{figure}

Figure~\ref{fig:DeepFilterNet} shows a schematic depiction of the DeepFilterNet noise reduction framework.
The model operates in two stage: 1.~Enhancing the speech envelope at a coarse frequency resolution, and 2.~enhancing the speech periodicity using DF.

DeepFilterNet operates in \SI{48}{kHz} sampling rate on \SI{20}{\ms} windows with a hop size of \SI{10}{\ms}.
Additionally, a look-ahead of 2 frames is used resulting in an overall latency of \SI{40}{\ms}.
The first stage operates in real-valued ERB (equivalent rectangular bandwidth) domain and predicts 32 ERB scaled gains that are pointwise multiplied with the noisy spectrum, which allows to reconstruct the overall speech envelope.
The second stage predicts a complex filter $N=5$ tap filter in frequency domain.
This filter is only applied for the lowest 96 frequency bins, i.e.~up to \SI{4.8}{\kHz}.
The final enhanced output spectrum is then constructed from DF output for the lower frequencies and the ERB gain output for the higher frequencies.

We take advantage of several properties of speech production model and psychoacoustics:
\begin{itemize}
  \item The speech consists of a noisy and a periodic (short-time correlated) component \cite{quatieri2002discrete}.
  \item Loudness perception is logarithmic.
    This is used in the ERB feature calculation and the loss.
      \item Frequency perception is logarithmic.
    ERB feature logarithmically compress the input spectrum with 481 frequency bins to 32 ERB bands which reduces dimensions of the neural network input and output~\cite{moore1988comparison}.
  \item Most of voiced (periodic) speech energy is below \SI{5}{\kHz}.
  \item Moreover, the human ear is most sensitive around \SIrange{2}{5}{\kHz}~\cite{suzuki2004equal}.
    Thus, the second stage enhancing the speech periodicity via DF is only applied up to approx.~\SI{5}{\kHz}.
    This further reduces input and output dimension of the neural network.
\end{itemize}

We further predict the local SNR $\xi \in [-15, 35]$~\si{\dB} within the encoder network on frame level.
This allows to completely disable the ERB or the DF decoder depending on the current noise conditions.
That is, we define the following criteria:
\begin{enumerate}[leftmargin=2cm]
  \item[$\xi< \SI{-10}{\dB}$:] Disable both decoders, return silent spectrum.
  \item[$\xi> \SI{20}{\dB}$:] Disable DF decoder. Only low noise condition, enhancing the periodicity is not necessary.
  \item[else:] Run all stages for best noise reduction.
\end{enumerate}

We implemented a real-time loop in Rust and use tract\footnote{\url{https://github.com/sonos/tract}} as DNN inference framework.
DeepFilterNet achieves a single-threaded real-time factor of 0.19 on a notebook i5-8250U CPU and is published under a permissive license\footnote{\url{https://github.com/Rikorose/DeepFilterNet}}.

\vspace{-.3em}
\section{Experiments and Results}
\vspace{-.2em}
\label{sec:results}

We train the slightly modified DeepFilterNet model on the full multi-lingual DNS4 dataset~\cite{dubey2022icassp}, while oversampling the high-quality PTDB and VCTK datasets by a factor of 10 and evaluate on the unseen VCTK/DEMAND test set (Table~\ref{tab:vctk_demand}).
\begin{table}[bh]
  \caption{Objective results on Voicebank+Demand test set}
  \vspace{-.7em}
  \label{tab:vctk_demand}
  \robustify\bfseries
  \sisetup{
    table-number-alignment = center,
    table-figures-integer  = 1,
    table-figures-decimal  = 2,
    table-auto-round = true,
    detect-weight = true
  }
  \resizebox{\linewidth}{!} {
    \begin{tabular}{
        l S S S S S[table-figures-decimal=3]
      }%
      \toprule%
      Model &\text{PESQ} & CSIG  & CBAK & COVL & STOI \\%
      DeepFilterNet \cite{schroeter2022deepfilternet}   & 2.81 & 4.14 & 3.31 & 3.46 & 0.942 \\%
      DeepFilterNet2 \cite{schroeter2022deepfilternet2} & 3.08 & 4.30 & 3.40 & 3.699& 0.9429 \\%
      DeepFilterNet3                                    & \bfseries3.17 & \bfseries4.34 & \bfseries3.61 & \bfseries3.77& \bfseries0.944 \\%
    \end{tabular}
  }
  \vspace{-.7em}
\end{table}%

\begin{figure}
  \centering
  \includegraphics[width=.88\linewidth,trim=0.1cm 1.0cm 0 1.2cm, clip]{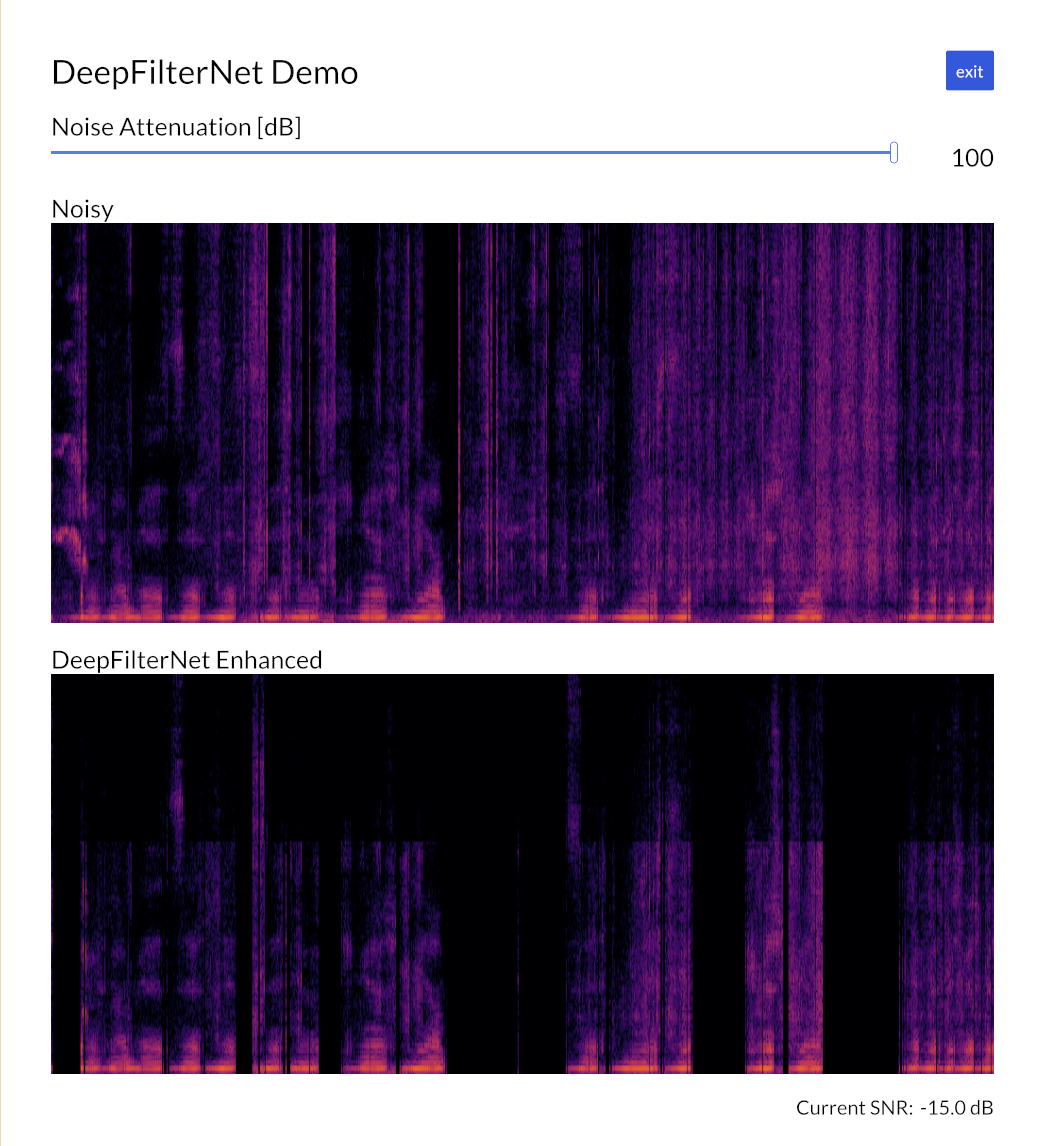}
  \caption{DeepFilterNet Live Demo.}
  \vspace{-.9em}
  \label{fig:demo}
\end{figure}

Figure~\ref{fig:demo} shows the proposed demo for live background noise suppression.
Due to its efficiency, we can listen to the audio in real-time.
Further, we can dynamically configure the noise attenuation and modify thresholds where the first and second denoising stage are disabled.
Moreover, the DeepFilterNet real-time implementation can also be used for live noise reduction by adding a virtual microphone on Linux systems with pipewire via an LADSPA plugin.
This allows for real-time noise reduction e.g.~during video calls.

\bibliographystyle{IEEEtran}
\bibliography{refs}

\end{document}